\documentclass[a4paper,20pt]{article}
    \usepackage[T1]{fontenc}
    \usepackage{graphicx}
    \usepackage{epsfig}
    \usepackage{floatflt}
    \usepackage{amsmath}
    \usepackage{amsfonts}
    \renewcommand{\abstract}{}
    \def\kms{km\thinspace s$^{-1}$}
\begin{document}
\makeatletter
\renewcommand{\@oddhead}{\textit{YSC'14 Proceedings of Contributed Papers} \hfil \textit{S.L. Malchenko}}
\renewcommand{\@evenfoot}{\hfil \thepage \hfil}
\renewcommand{\@oddfoot}{\hfil \thepage \hfil}
\fontsize{11}{11} \selectfont

\title{Duplicity and Evolution Status of the Early-Type Be Star V622\,Per, the Member of the $\chi$~Per Open Star Cluster}
\author{\textsl{S.L. Malchenko}}
\date{}
\maketitle
\begin{center} {\small Tavrida National University, Simferopol, Crimea, Ukraine\\ Svetlana\underline{ }mal\underline{ }81@mail.ru}
\end{center}

\begin{abstract}
Radial velocities analysis based on high-resolution spectra,
obtained in the H$\alpha$ region and low resolution spectra obtained
in the region 4420-4960\AA\, together with radial velocities, taken
from other published sources allow us to calculate orbital
parameters of the massive binary system V622 Per. It is shown that
the system has an orbital period 5.214(29) days, T$_0$ = 2450661(4)
and is a post mass transfer binary. From light curve analysis of the
ellipsoidal variability we obtained inclination angle of the system
and temperature of the components. Luminosity ration of the
components was found of about 4:1. T$_{eff}$ and $log\,g$ were
estimated for each component. It is shown that primary, less massive
but more bright star, is an evolved object that has lost large part
of its mass during the evolution. Estimations of chemical
composition of the primary show noticeable enrichment by products of
the CNO cycles. E.g. He/H reaches 0.18, nitrogen is in excess of
about 0.5~dex, carbon has low abundances (by 2-3 dex lower) and
oxygen has 1 dex lower than solar abundance. The possible evolution
of the binary with the known age 14 Myrs is discussed.
\end{abstract}

\section*{Introduction}
\indent \indent Theory of the stellar evolution provides a
description of the interior structure and the observable properties
of a star given initial mass and chemical composition as a function
of age. Significant progress has been made during recent years in
understanding the physical processes that govern the structure and
evolution of stars. Evolutional models are improving and the main
direction of modeling is to account for evolution parameters of
conservation system on the one hand and exchange by angular momentum
on the other hand.

The close binaries are powerful tools for testing stellar structure
and evolution models, since the fundamental properties of the
components (masses, radii, luminosities, etc.) can be accurately
determined from the observations. These systems in young open
clusters provide a way of finding the age, distance, accurate
masses, radii and chemical composition, making a good discriminating
test of the physical ingredients of theoretical models. Presences of
massive interacting binaries in open stellar clusters is the useful
tools for understanding of short-lived phases of their evolution.
Such stars are rare and each of them requires detail analysis.

During the studying of the hot B stars in young open clusters
h/$\chi$~Per we found and analyzed binary system V622~Per. The star
may be a good indicator to verify the theories of evolution. Early
spectral type of the star B2III \cite{Strom et al. 2005}, relatively
short orbital period $\sim$5.2$^{d}$ \cite{Pigulski}, presence of
emission details in the spectrum and unusual chemical composition of
the atmosphere \cite{W} mean that the star is an interacting binary
with unknown evolutional status, but located in the open stellar
cluster $\chi$~Per of the known age.

From available data as well as those given in literature we have
calculated orbit, period, temperature, gravity, rotational
velocities. We also have estimated chemical composition of the
components.

\section*{Observations and data reduction}
\indent \indent The spectroscopic observations of V622~Per were
carried out over four years from 1997 to 2000 as a part of studying
emission spectrum of the Be stars in the young double open cluster h
and $\chi$\,Per (NGC 869 and NGC 884). We used the Coud\'e focus of
the 2.6-m telescope of the Crimean Astrophysical Observatory. The
spectral resolution was about 30000. The signal-to-noise ratio was
$\sim$100. A total of 8 spectra were obtained in the H$\alpha$ line
region and one in the region of the HeI\,$\lambda$\,6678\,\AA\,
line.

Additionally, as a part of program of studying B and Be stars in the
open stellar clusters, two medium 2.5\,\AA\, resolution spectra were
obtained in the Nasmyth focus of the same telescope. They cover
spectral region between 4420-4960\,\AA. The signal-to-noise ratio of
these spectra was about 100.

\begin{floatingfigure}[lfrt]{80mm}
\includegraphics[width=70mm,height=90mm]{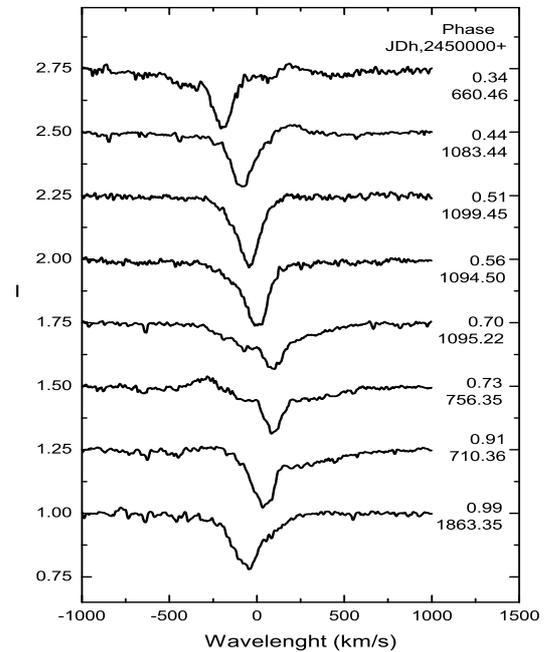}
\caption{\small H$\alpha$ line profiles of V622 Per. On the right
side of each spectrum JD and phase of the orbital period is
presented. Intensities of each next spectrum are shifted by 0.25.}
    \label{fig.1.}
\end{floatingfigure}
\hspace{8cm}

\begin{table}[h]
        \caption{\small{Orbital parameters of V622\,Per based
        \hspace{140mm}
                         on on radial velocity variability}}
        \label{tab1}\vspace{5mm}
    \begin{tabular}{cc}
\hline
    \small Element                           & \small Orbital solution  \\
    \hline
    $P$ (days)                        & 5.21429$\pm$0.00008  \\
    $T_{conj.1}$                      & 2450661.4$\pm$0.2  \\
    $K_{1}$ (\kms)                    & 139$\pm$6  \\
    $K_{2}$ (\kms)                    & 99$\pm$11 \\
    $q$                               & 1.40$\pm$0.13 \\
    $e$                               & 0.05$\pm$0.04 \\
    $\omega^\circ$                    & 236$\pm$36 \\
    $\gamma_{1}$ (\kms)               & -44$\pm$3 \\
    $\gamma_{2}$ (\kms)               &  12$\pm$11      \\
    $f_{M}$ (M$_{\odot}$)                 & 1.46         \\
    $M_{1}sin^{3}i$ (M$_{\odot}$)      & 3.0     \\
    $M_{2}sin^{3}i$ (M$_{\odot}$)      & 4.3     \\
    $a_{1}sin\,i$ ($R_{\odot}$)        & 14.3     \\
    $a_{2}sin\,i$ ($R_{\odot}$)        & 10.2     \\
    No. of spectra                    & 11 spectrograms\\
                                                                        & and    3 velocities from \cite{Liua,Liub}      \\
        \hline
    \end{tabular}
    \end{table}
\vspace{17mm}

Radial velocity (RV) and equivalent width (EW) of the eight
H$\alpha$, one HeI\,$\lambda$\,6678\,\AA\, and two medium resolution
Nasmyth spectra were obtained. As it is seen from Fig.1, H$\alpha$
line has complex profile and it is variable in the time domain. The
most pronounced component is a sharp absorption line with large
amplitude of RV variation. The signatures of the broad absorption
component are also seen on the most part of our spectra.
Additionally, some faint emission line is presented in the red or
blue wings of the line, but some spectra have no noticeable emission
or emission is hidden inside of the absorption profile.

The He\,I\,$\lambda$\,6678\,\AA\, line profile can be seen in Fig.2.
It has single-component line profile without noticeable emission and
some faint signatures of the additional absorption component in the
blue wing of the line.

The observed blue region of the spectra is presented in Fig.4. The
H$\beta$ line profile has no signs of emission component on the both
of our spectra. The weak absorption from the secondary is seen in
the blue wing of the line. As it seen from Fig.4, the red wing of
the H$\beta$ line has broad depression. It was found on all the
spectra of other members of h/$\chi$~Per cluster, but absent in the
spectra of the standard stars from the list in work \cite{Lybimkov}.
It is associated with the interstellar absorption band with unknown
identification \cite{Herbig} and it has a very broad, asymmetric
feature. According to \cite{Herbig} the wings of the line extend
shortward edge to at least 4870\,\AA\, and longward to 4909\,\AA\,.
The deepest point is about 4882\,\AA\,.

\section*{Radial velocities analysis and orbital solution}
\indent \indent According to the rich BV photometry from the paper
\cite{Pigulski}, V622\,Per is an ellipsoidal double system with the
orbital period $P_{orb}$ = 5.2$^{d}$. The large fraction of our RV
measurements were obtained from the emission H$\alpha$ line.
Practically the same velocities of the He\,I and sharp component of
the H$\alpha$ lines, obtained in the same night, and "in phase"
variability of the radial velocities, obtained from the Nasmyth
spectra, allow us to conclude that sharp component of the H$\alpha$
line mostly appears in the photosphere of the bright star and can be
with some caution used in solving orbit of V622\,Per. To confirm the
value of orbital period derived in \cite{Pigulski} we used our RV
measurements together with the data, obtained in \cite{Liua,Liub}.
Periodogram analyses based on nonparametric statistics
(Lafler-Kinman) were used for searching possible orbital period from
radial velocities observations. Only one significant period, close
to the value proposed in the work \cite{Pigulski}, was found.

In order to solve spectroscopic orbit we used the FOTEL code
\cite{Hadrava 1990}. Obtained orbit solution is presented in the
Table~1 and its graphical equivalent is present in Fig.3. As it is
seen from orbital solution, V622\,Per is an evolved massive system
with the less massive, but more bright primary component. It has
near circular orbit and low value of mass exchange.

\begin{figure}[h]
\begin{minipage}[b]{.45\linewidth}
\centering\psfig{figure=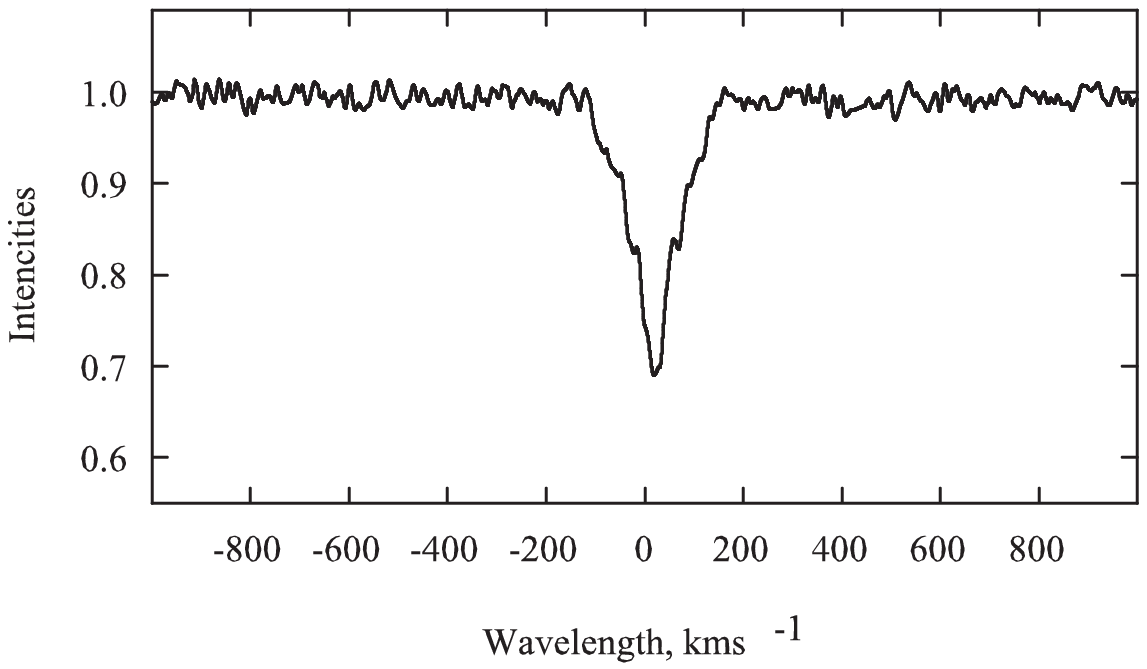,width=70mm,height=45mm}
\caption{\small{Line profile of the He\,I\,$\lambda$\,6678\,\AA\,
line, obtained at HJD = 2451094.577} \vspace{27mm}}
\end{minipage}\hfill
\begin{minipage}[b]{.45\linewidth}
\centering\psfig{figure=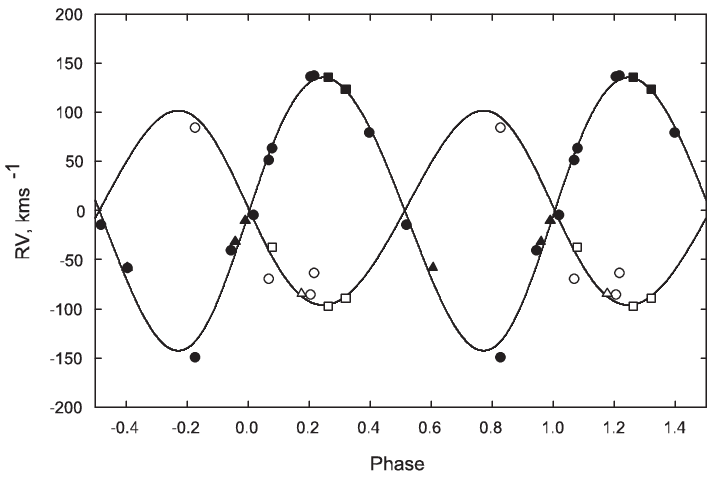,width=70mm,height=40mm}
 \caption{\small{Radial velocity variability with the phase of orbital period. The filled symbols -- orbital velocities of the primary component, open symbols -- RVs of the secondary
component. Filled circles -- sharp component of the H$\alpha$ and
the He\,I\,$\lambda$\,6678\,\AA\, lines. Circles -- RVs derived from
the H$\alpha$ line profile; squares -- RVs, obtained from the two
Nasmyth spectra and the He\,I\,$\lambda$\,6678\,\AA\, line;
triangles -- RVs estimations from \cite{Liua,Liub}} open triangle --
ommited observation \cite{Liub}}
\end{minipage}\hfill
\end{figure}

\begin{figure}[!b]
\includegraphics[width=160mm,height=80mm]{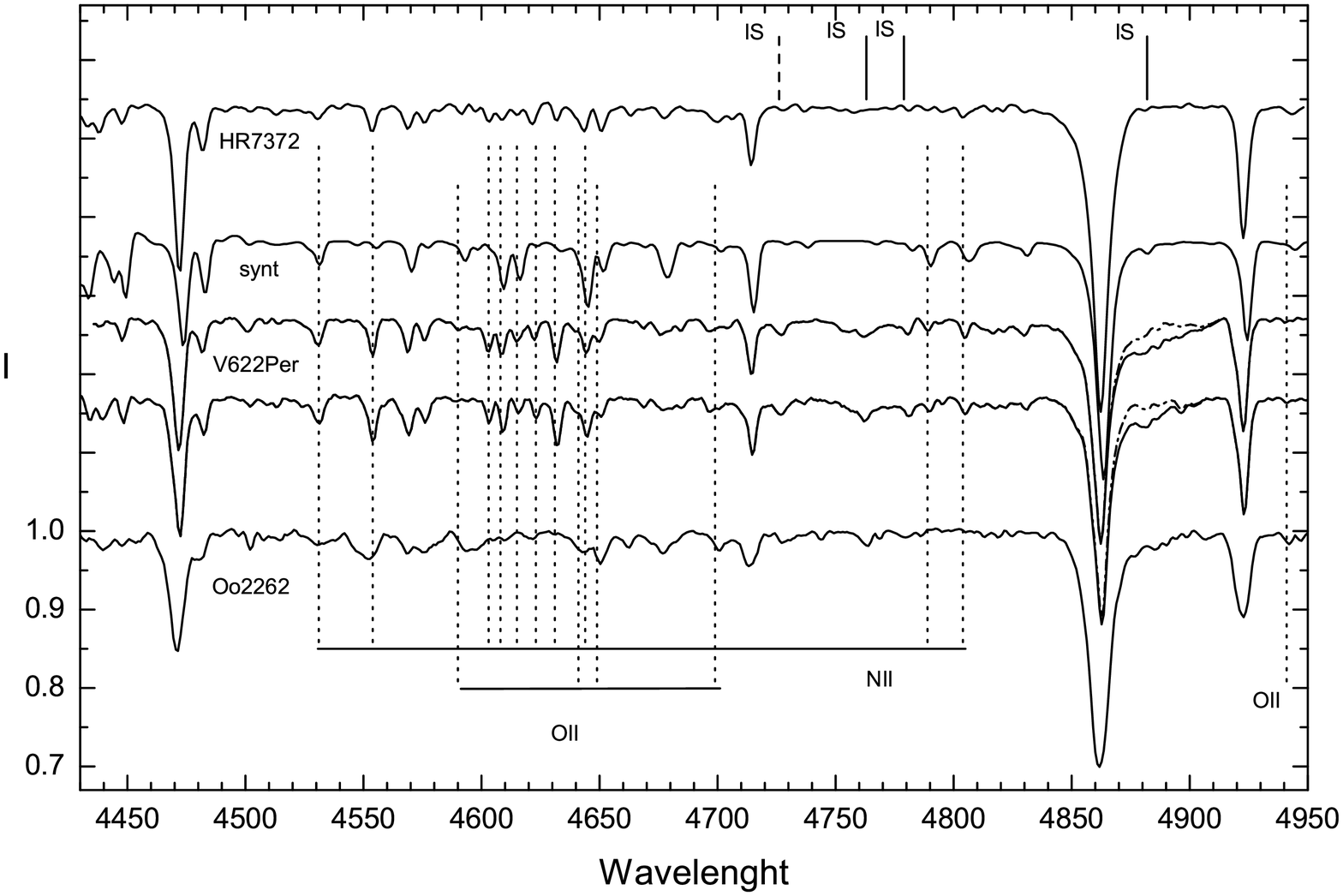}
\caption{\small{Medium resolution spectra of V622 Per, obtained in the spectral region 4420-4960 \AA\, together with the spectra of two comparison stars HR6787 from list in \cite{Lybimkov} and member of $\chi$ Per cluster Oo2262. The calculated synthetic spectrum is also shown. Positions of stellar spectral lines with estimated abundances is present together with interstellar lines (IS). The dashed lines in the region of the blue wing of the H$\beta$ line are removed interstellar band.}}
\label{fig3}
\end{figure}

\section*{Physical parameters and chemical composition of the components}
\indent \indent We also used BV photometry data from \cite{Pigulski}
to perform light curve analysis. According to calculation by Yakut
(private communication) V622\,Per is an ellipsoidal double system,
with colder primary component ($T_1$= 21000K) and hotter secondary
($T_2$= 24000K). Inclination angle of the system i =
$43^{\circ}.7\pm2^{\circ}.9$ was found by Yakut.

Radial velocities and photometrical variability due to
ellipsoidality of the components allowed us to obtain most of the
main physical parameters of the double system with exception of the
radius of the components. The next step of our analysis was to
constrain a model of atmosphere of the components with the goal to
estimate chemical composition at least of more luminous star.

The temperature $T_{eff}$ of the components were found from the
light curve analysis. Next pair of parameters $log\,g_{1}$ and
$log\,g_{2}$ of the components  can be found from the equivalent
width of the H$\beta$ line, but only in assumption that gravity one
of the components is taken elsewhere. We had to accept that the less
luminous component is still an  undeveloped star whose position on
H-R diagram is close to the main sequence with $log\,g=4.0$. Then,
using luminosity ratio of the components 4:1 and photometry index
[c1] and $\beta$ (took from \cite{Capilla and Fabregat} and
\cite{Fabregat}), we found that observed EW of the H$\beta$ line had
satisfied approximation with $log\,g_{1}=3.0\pm0.5$.

The last step of our analyses was the determination of the
abundances of the elements of CNO cycle. We used the LTE line
blanketing model from \cite{Kurucz} for solar abundances with the
line formation problem solved by \cite{Tsimbal} in the program SyntV
for finding the basic parameters of the atmosphere of the cool
component of V622~Per and estimation of the chemical composition.

The synthetic spectra were calculate with the parameters of
atmosphere each of the component taken from our data ($T_1$ =
21000K, $log\,g_{1}$= 3.0, $T_2$ = 24000K $log\,g_{2}$= 4.0,
$Vsin\,i$= 60\,\kms, $V_{turb}$= 10\,\kms\,). From analysis of
synthetic spectrum we have found that CNO abundances of V622~Per are
far from solar. Nitrogen lines demonstrate overabundance in
comparing to the normal solar abundance. Oxygen abundance is
noticeable lower to solar. And even on height-resolution spectra in
the H$\alpha$ region we can not see presence of the C\,II\, doublet
at the $\lambda$6578\,\AA\, and $\lambda$6582\,\AA\,. The deficiency
of carbon is presented in the atmosphere of the both of components
and should be at least 2-3 dex.

Quality of our data allowed us to obtain only estimations of
chemical composition. Abundance of helium He/H is 0.18, excess of
nitrogen is $\sim0.5$ dex and deficiency of oxygen are about 1 dex
in comparison to solar abundances.

The chemical composition of V622~Per is near similar with the
composition of $\beta$ Lyr - well know massive interacting binary
with \textit{P$_{orb}$}= 12.9 days. In work \cite{Balachandran}
obtained large He enrichment, extreme nitrogen overabundant and very
underabundant oxygen and carbon. Carbon lines, the same as in the
case of V622 Per, were not found in the atmosphere of the primary
component.

\section*{Conclusions}
\indent \indent Relatively short orbital period after active mass
exchange in V622Per allow us to conclude that it is type a massive
interacting binary with the masses of the component $M_1$=
9.0M$_{\odot}$ and $M_2$= 12.8M$_{\odot}$\,. The less massive
evolved primary leave main sequence and it is on the way to the red
giants stars.

Thereby we have researched massive binary system V622~Per with well
know age at 14 million year, V = 9.25, period 5.213$^{d}$\,. We
found temperature angular momentum, gravity, masses and  estimated
the chemical composition of the first component.

\section*{Acknowledgements}
\indent \indent I am grateful to A.E. Tarasov for help, comments and
suggestion during the development of this paper. I am also grateful
to Dr. K. Yakut for light curve analysis and Dr. D.V. Shulyak for
help in calculating of model atmosphere.

\end{document}